# Ferro-hydrodynamics of droplet necking filaments

Neeladri Sekhar Bera,[1] Apurba Roy,[1] and Purbarun Dhar[1]
*Hydrodynamics and Thermal Multiphysics Lab (HTML), Department of Mechanical Engineering, Indian Institute of Technology Kharagpur, West Bengal-721302, India*

(*Electronic mail: purbarun@mech.iitkgp.ac.in ; neeladrisekharb@gmail.com)

(Dated: 21 December 2025)

We explore the necking, filament thinning, and pinch-off dynamics of ferrofluid droplets within a magnetic field, via a simple and low-cost experimental method. In our studies, both the Ohnesorge number (*Oh*) and the Deborah number (*De*) are $O(1)$: a typically inaccessible regime with conventional extensional rheometers. Under magnetic forcing, the nanoparticles assemble into field-aligned, chain-like structures, that generate a tunable magneto-elastic response, and markedly alter the extensional flow. Although behaving as Newtonian liquids in the absence of a magnetic field, the field induces extensional thickening, and the emergence of beads-on-a-string (BOAS) structures in the ferrofluid filaments: a non-Newtonian signature. By combining controlled elongation with high-speed imaging, we directly quantify the magnetic-field-dependent extensional viscosity and relaxation time. Our findings underscore how magnetically induced microstructures govern filament stability and extensional dynamics in ferrofluids.

Ferrofluids (FF) respond strongly to magnetic fields (MF) by virtue of magnetic nanoparticles (MNP) suspended stably in a carrier liquid. Since inception and development, ferrofluids have attracted sustained academic and research interest since they juxtapose the ability to deform and flow like ordinary liquids, with the possibility of being magnetically controlled or actuated.[1,2] This combination has birthed several applications, *viz.* in rotary seals and bearings, vibration damping and acoustic devices, magnetically assisted heat transfer systems[3], and adaptive fluid-based technologies[4–6], etc. In recent years, they have made a strong presence in biomedical avenues, such as magnetic drug delivery[7,8], hyperthermia treatment[9], and contrast-imaging, wherein precise control of fluid deformation, dosage, and transport are crucial[10,11]. Many such utilities involve droplet formation and/or filament stretching within MF. Therefore, understanding the rheology of FF during magnetically induced extensional flows holds crucial importance.

Rheological characterization typically comprises shear and extensional rheology (ER). Shear rheology probes the response of a fluid to tangential deformation, and has been used widely for FF, particularly to quantify magnetoviscous effects arising from field-induced particle chains and anisotropy[2]. ER, on the contrary, focuses on how a fluid responds when stretched, a mode of deformation that is paramount to droplet formation, necking, filament thinning, and breakup. Even for simple Newtonian fluids, these processes result from a delicate balance between surface tension, viscosity, inertia, and gravity, leading to rich and often universal dynamics near pinch-off[12–15]. When a fluid has internal structure or elasticity, extensional flows become exceedingly informative, as stretching morphs microstructural alignment and elastic stresses, strongly influencing necking, filament dynamics, and breakup behavior[16–18]. In the recent decade, experimental techniques such as **Fi**lament **S**tretching **E**xtensional **R**heometry (*FiSER*) and **Ca**pillary **B**reakup **E**xtensional **R**heometry (*CaBER*) have been developed to quantify extensional viscosity and relaxation times, with successful application to a wide range of complex fluids[19–22].

Extensional behavior of FF develops further complexities in a MF, as magnetic interactions promote nanoparticle-chain-formation and anisotropic structures, leading to non-Newtonian behavior, and altering the stresses borne of stretching. Despite a body of literature on magnetically induced changes in shear rheology, experimental studies of magneto-ER are still number in few. Early research probed magnetic colloids using extensional flows, and showed elongational viscosity as strongly sensitive to MF strength[23]. Research has remained concentrated to configurations where the MF is aligned with the primary deformation direction, and capillary breakup extensional magneto-rheometry has been proposed more recently to quantify magnetically induced extensional stresses[24]. In most literature, MF are aligned with either the direction of stretching, or with gravity. Additional non-trivial physical mechanisms are introduced when transverse magnetic stresses interact with elongational deformation and with the changing microstructure in extensional flows under a transversely applied MF. Such configurations remain scarcely explored in experiments, despite their practical significance.

For dilute FF, where the Ohnesorge number ($Oh = \eta_f / \sqrt{\rho_f \sigma_f L_c}$) and Deborah number ($De = \lambda_E \sqrt{\sigma_f / \rho_f L_c^3}$) are $<1$ (refer in fig. 2 of the paper[25]), the quantitative measurements of both extensional viscosity ($\eta_e$) and extensional relaxation time ($\lambda_E$) lie beyond the capability of commercially available devices like *CaBER*. Here, $\eta_f$, $\rho_f$ and $L_c$ denote the dynamic viscosity, and density of the FF, and characteristic length-scale, respectively. Later, by obtaining the typical ranges of these numbers, we can confirm the utility of the present experimental method. Another problem is that strong and uniform MF oriented horizontally with respect to the filament axis are not specifically accommodated for by conventional ER systems like *CaBER* and *FiSER*. Measurement accuracy is largely limited by issues with magnet positioning, optical access, and MF homogeneity during rapid filament thinning. Consequently, important parameters like extensional viscosity, relaxation times, and necking dynamics of FF under horizontal magnetic fields remain mostly unknown.

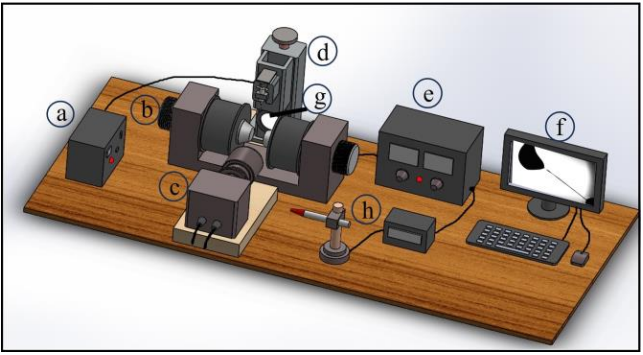

FIG. 1: Schematic of the experimental setup (a) Droplet dispensing system controller, (b) Electromagnet, (c) High speed camera, (d) Digitized micro-droplet dispenser, (e) Programmable power supply for electromagnet, (f) Computer system for camera control and data acquisition, (g) Continuous light source, and (h) Gaussmeter

In this letter, we address this gap and discuss a dedicated experimental approach to study the magneto-ER of vertical FF filaments, subjected to a horizontal MF. Combining controlled elongation deformation[26] within a uniform transverse MF, and high-speed imaging of filament thinning, we directly measured the field-dependent $\eta_e$ and $\lambda_E$. These results provide new insights into how MF orientation affects neck evolution, filament stability, and microstructural dynamics during extensional flow, and are relevant to both fundamental studies and applications involving magnetic fluid systems.

The experimental setup is shown schematically in Fig. 1. FF droplets were dispensed using a precision micro-syringe fit with a 22-gauge needle, using a digitized microdroplet dispenser system (Holmarc, India). The MF was generated between the poles of an electromagnet (Holmarc, India), powered by a programmable power supply (Polytronic Corp., India). The MF strength at the center of the pole gap was measured and calibrated using a GaAs sensor–based Hall-effect gaussmeter (Holmarc, India) by varying the input current. We have varied the MF strength between 0-0.3 T. The needle was carefully aligned to dispense droplets exactly at the center of the pole gap (maintained at 22 mm throughout). The fluid dynamic events were captured using a high-speed camera (Photron, UK) equipped with a 105 mm macro lens (Nikon). Images (resolution of 1024 × 1024 pixels) were recorded at 8000 frames per second .

Stable FF were prepared by dispersing $Fe_2O_3$ nanoparticles (30-40 nm diameter, Alfa Aesar, India) in base fluid (84 vol.% glycerol and 16 vol.% deionized (DI) water). Different concentration samples, ranging from 2 to 15 wt.%, were prepared. The nanoparticles were coated with citric acid (as supplied by the manufacturer), which enhances the colloidal stability. The FF were first mechanically stirred for 30 min and then ultrasonicated for 1 h by probe-type sonicator (Oscar Ultrasonics, India). Zeta potential (Malvern Instruments, USA) values lay outside the range of -30 mV to +30 mV, indicating sufficient colloidal stability. FF were found to be stable over timescales significantly longer than those of the experiments. The surface tension of the FF was measured using the pendant-drop method and image analysis (using ImageJ). Neck and filament diameter measurements were initiated when the neck diameter became equal to the outer diameter of the 22G needle (0.70 mm), as the needle was slightly wetted by the FF. The physical properties of the different FF are shown in Table I. All experiments were conducted at a controlled temperature of 22°C ± 1°C. The variation in properties and measurements of ferrofluid droplets generated is noted within ±5%. Each set of experiments has been repeated thrice.

TABLE I: Physical properties of the $Fe_2O_3$ FF studied.

| wt.% of $Fe_2O_3$ | $\rho_{particle}$ (kg/m$^3$) | $\sigma_{ferrofluid}$ (mN/m) |
|---|---|---|
| 5% | 1061.5 | 67 ± 2 |
| 10% | 1061.5 | 66 ± 1 |
| 15% | 1061.5 | 64 ± 1 |

The temporal evolution of the minimum neck diameter for FF of different concentrations is illustrated in Fig. 2. The pinch-off time increases consistently with increasing particle loading (at a constant MF strength). This pattern suggests that the thinning dynamics gradually slow down with increasing concentration. More $Fe_2O_3$ increases the inter-nanoparticle magnetic interactions, encouraging the formation of more robust and long-sustained chain-like structures that are in line with the applied field. By increasing the FF's extensional magnetoviscosity in such manner, the field-induced structures decelerates the elongation deformation. Consequently, at higher concentrations, the filament life-time is prolonged.

Figure 3 illustrates the temporal evolution of the minimum neck diameter for increasing values of the magnetic Bond number, $Bo_m$, where,

$$Bo_m = \frac{(1+\chi)^2 B^2 L_c}{\mu_0 \sigma_f}. \quad (1)$$

It quantifies the competition between magnetic and capillary forces in the FF. Here, $\chi$ is magnetic susceptibility, $B$ is magnetic flux density, $L_c$ is characteristic length scale (here, the outer diameter of the dispensing needle), $\mu_0$ is the magnetic permeability of free space, and $\sigma_f$ is surface tension of the FF. The values of $\chi$ correspond to saturation magnetization and are adopted from Rigoni et al.[27]. In the absence of MF, the filament exhibits the longest pinch-off time. Upon application of a weak MF (low $Bo_m$), magnetic nanoparticles begin to form chain-like structures aligned with the field lines, exhibiting a reduction in filament lifetime. As $Bo_m$ is further increased, these chains become robust and persistent, resulting in an increase in the extensional magnetoviscosity, and a corresponding delay in the pinch-off process. However, beyond a critical value ($Bo_m > 525$), Joule heating associated with the electromagnet poles becomes significant, reducing the effec-tive viscosity of the filament and thereby decreasing the lifetime. In addition, Fig. 3b shows that the thinning dynamics undergo a qualitative change at higher MF strengths, with the minimum neck diameter $D_{min}$ exhibiting an exponential decay in time. This transition reflects the emergence of certain degree of non-Newtonian[28,29] behavior in the FF induced by the applied MF.

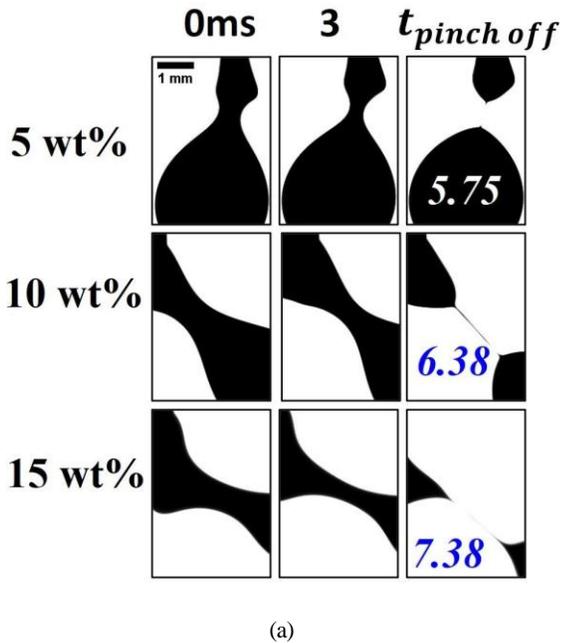
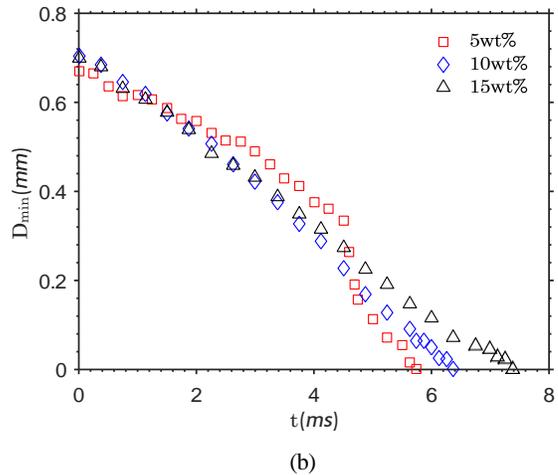

FIG. 2: The evolution of neck diameter $D_{min}$ with time for different FF concentrations. Here (a) illustrates an array of different time instances of pinch-off and (b) shows the evolution of $D_{min}$ with time. Here, all cases pertain to MF of 0.1 T.

To confirm the Newtonian nature of the FF in the absence of MF, we compare the evolution of the minimum neck diameter for different concentrations FF with the universal similarity solution proposed by Eggers[12]. The $Oh$ is found to be within $O(1)$ for all concentrations, typically $Oh \approx 0.4$–$0.6$. For a Newtonian fluid, the similarity solution predicts

$$D_{\min} \sim (t_c - t)^n, \qquad (2)$$

with $n = 2/3$ in the inertia–capillary regime ($Oh \ll 1$). For the 5, 10, and 15 wt.% FF, the exponent (fitting the data) increases from approximately 0.5 to 0.61, to 0.66. Although the viscosity, and hence $Oh$, increases with particle concentration, the exponents are extracted over a finite time window prior to the final breakup. In this experimentally accessible regime, inertia and surface tension forces remain dominant, while viscous effects become important only at smaller length scales. The good agreement with the Eggers similarity solution confirms that, at zero MF, the FF behave effectively as Newtonian liquids during pinch-off. All the fit curves in fig. 4 conform to $R^2$ value $> 0.97$.

From Fig. 5, it is evident that the neck-thinning dynamics of FF droplet undergoes a qualitative change under an applied MF. Particularly, the evolution of the minimum neck diameter obeys an exponential decay, akin to the elastocapillary regime observed during the thinning of viscoelastic filaments[18]. Here, magnetic nanoparticles form chain-like structures aligned with the MF, giving rise to a magneto-elastic stress component that competes with capillary forces, and leads to an exponential thinning regime. For viscoelastic filament thinning described by the Oldroyd-B model,

$$D_{min} \sim \exp\left(-\frac{t}{3\lambda_E}\right), \qquad (3)$$

where $\lambda_E$ is the extensional relaxation time[30]. We deduce $\lambda_E$ by fitting this expression to the experimental data in Fig. 5(a)–(c), obtaining excellent agreement ($R^2 \geq 0.98$) for all $Bo_m$. The variation of $\lambda_E$ with $Bo_m$, shown in fig. 5(d), reveals that $\lambda_E$ initially increases with field strength due to enhanced magneto-elastic stresses from chain formation, but decreases beyond a critical $Bo_m$ as strong magnetic forces disrupt the chains and reduce the elastic response. For all concentrations, the measurements yield the range $\lambda_E \sim 0.5$–$1.2$ ms. A similar qualitative trend (not illustrated) is observed for the shear relaxation time, although its magnitude is approximately an order of magnitude larger. With the obtained $\lambda_E$ we calculate the Deborah number (defined earlier) and get $De \sim O(1)$ (typically from 0.5 - 1.3). These ranges of $Oh$ and $De$ confirms the impracticality of the use of conventional extensional rheometers for our study.

Next, to quantify the extensional rheology of FF under the effect of MF, we define the apparent extensional viscosity $\eta_e$,

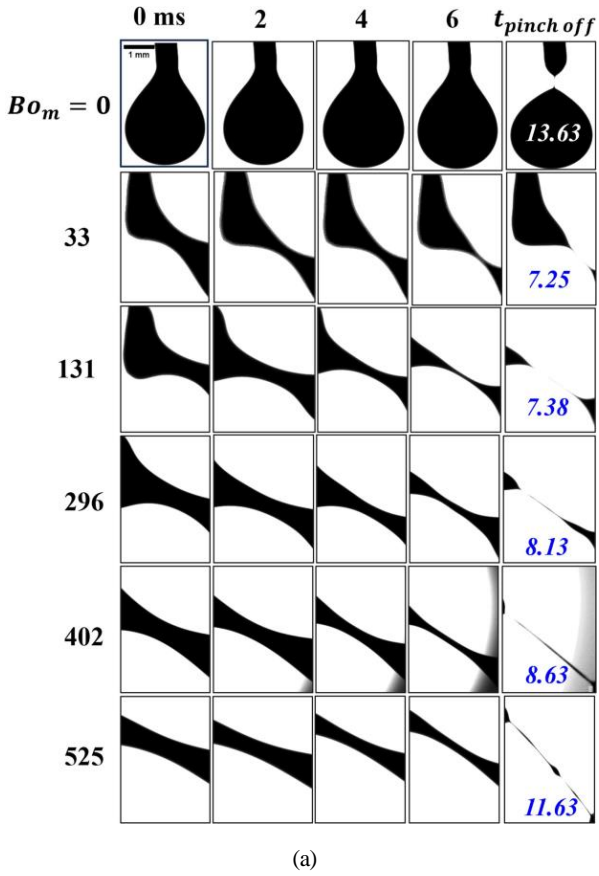

FIG. 3: The temporal evolution of neck diameter $D_{min}$ with increasing MF. (a) shows the time-snap array of the evolution of $D_{min}$ till pinch off, and (b) shows the measured $D_{min}$ with time with increasing $Bo_m$ for 15 wt.% FF. In (a), BOAS like structures, indicating certain degree of non-Newtonian behavior, are noted at higher $Bo_m$ (see 6th row, 5th column).

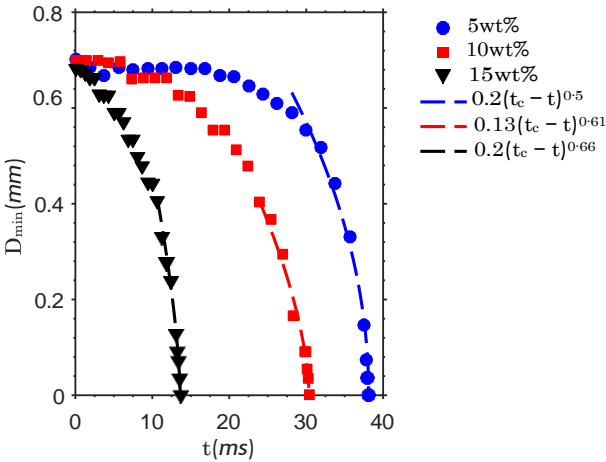

FIG. 4: Evolution of neck diameter for different concentrations of FF for MF strength = 0 T

and Hencky strain $\varepsilon_H$ as

$$\eta_e = -\frac{\sigma_f}{dD_{min}/dt}, \quad (4a)$$

$$\varepsilon_H = 2\ln\frac{D_0}{D_{min}(t)}, \quad (4b)$$

where $D_0$ is the diameter of the neck when the filament commences thinning. Fig. 6 shows the apparent extensional viscosity's evolution with Hencky strain, and demonstrates a distinct change in flow response as the strength of the MF rises. The FF displays extensional thinning in the absence of a MF (Fig. 6 (a)), with a negative slope of the fitting curve. This behavior is consistent with a Newtonian-like base fluid, where the resistance to stretching gradually decreases due to microstructural rearrangements, or via weak inter-particle interactions. When a MF is applied (Figs. 6 (b)–(c)), the trend reverses, and the extensional viscosity increases with strain, signifying extensional thickening. The positive slope of the fitting lines indicate progressive strain hardening brought forth by the field-induced chain and column formation of MNP. This births magneto-elastic stresses that progressively resist capillary thinning. The slope decreases and the thickening response weakens at higher $Bo_m$ (Fig. 6 (d)). This implies that the strong magnetic pull disrupts or rearranges the chains during rapid stretching beyond a critical MF strength, preventing additional elastic stress accumulation. In precis, the slope of the fitting curves captures the competition between capillarity, magnetism-induced-elasticity, and microstructural stability across a range of MF strengths, and offers a compact, yet physically consistent measure of the balance between extensional thinning and thickening.

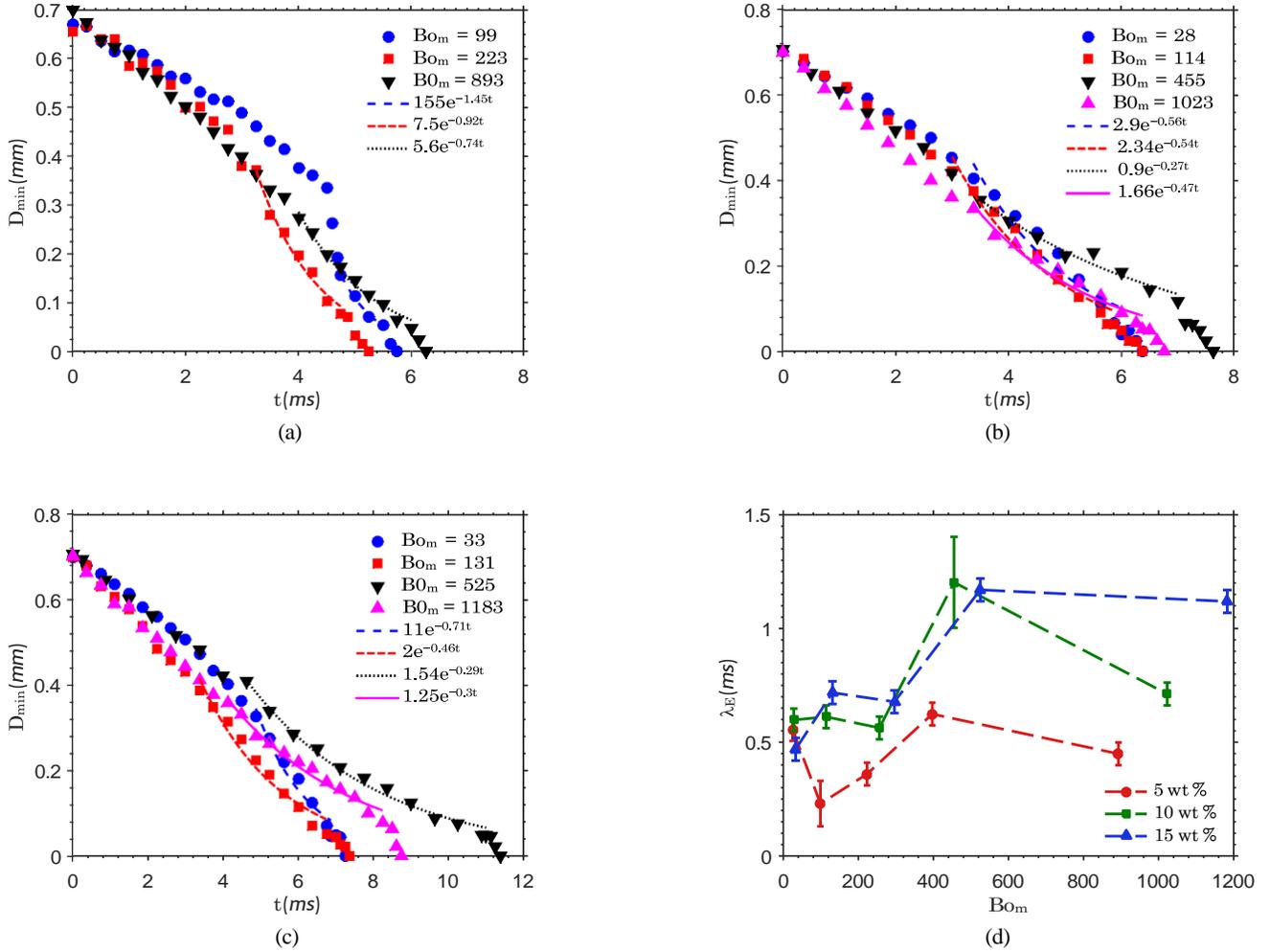

FIG. 5: Best-fit curves of the neck diameter evolution for FF with (a) 5 wt.%, (b) 10 wt.%, and (c) 15 wt.% concentrations at different $Bo_m$. Plot (d) shows the variation of $\lambda_E$, as a function of $Bo_m$ for the three FF concentrations.

Figure 7 summarizes the variant filament thinning mannerisms as function of MF strength and MNP concentration; broadly classified into three regimes. In *Regime I*, pinch-off occurs at the bottom of the filament, resembling conventional capillary breakup. *Regime II* is characterized by the appearance of BOAS, arising from the competition between capillary forces and magneto-elastic stresses induced by field-driven MNP chaining, suppressing uniform thinning and promoting bead formation. In *Regime III*, corresponding to high MNP concentration and strong MF, a long filament forms but pinch-off shifts toward the parent droplet-end. Here, the filament segment attached to the detaching drop is strongly stiffened by magneto-elastic stresses, while weaker magnetic resistance near the parent drop permits capillary forces to dominate, initiating filament breakup at the top.

To infer,, we experimentally investigate filament thinning and pinch-off in FF exposed to a transverse MF. The findings show that a tunable magneto-elastic response, that significantly alters extensional flow dynamics, is introduced by MF-induced MNP chaining. Within MF, the FF exhibit extensional thickening, BOAS formation, and field-dependent extensional relaxation times, despite their effective behavior as Newtonian liquids at MF=0 T. A regime map delineates different breakup modes, such as a shift to top-pinch-off at high MF strengths and MNP concentrations; controlled by the competition between capillary and magneto-elastic stresses. Our results show that transverse MF can be a useful control parameter for extensional rheology of FF.

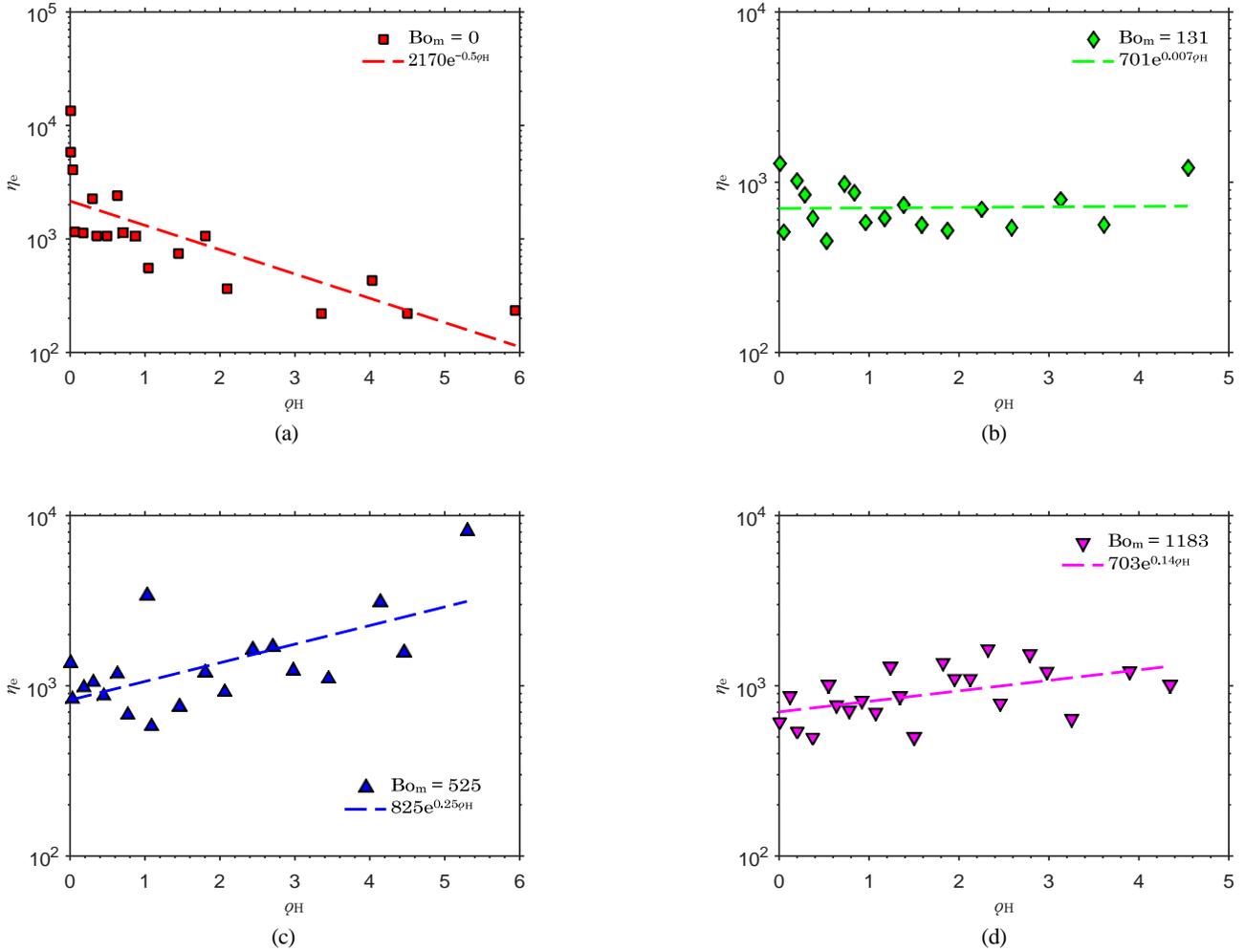

FIG. 6: Variation of extensional viscosity $\eta_e$ (mPa-s) with Hencky strain $\varepsilon_H$ for (a) $Bo_m = 0$, (b) $Bo_m = 131$, (c) $Bo_m = 525$, and (d) $Bo_m = 1183$, for a 15 wt.% FF.

**Conflict of Interest Statement** – The authors have no conflicts to disclose.

**Data Availability Statement** - The data that support the findings of this research are available in the article.

**Acknowledgment** - The authors thank the Central Power Research Institute (*CPRI*), India, and the Anusandhan National Research Foundation (*ANRF*), India, for funding the research.

**Author Contributions** - *NS Bera*: Conceptualization, Visualization, Validation, Methodology, Investigation, Formal analysis, Data curation, Writing – initial and final draft. *A Roy*: Methodology, Investigation, Data curation, Writing – initial draft. *P Dhar*: Conceptualization, Investigation, Methodology, Formal analysis, Supervision, Resources, Project administration, Funding, Writing – initial and final draft.

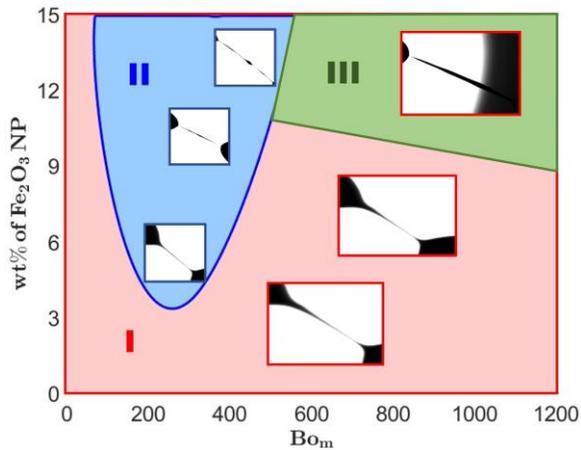

FIG. 7: Regime map illustrating different filament thinning behaviors as a function of magnetic field strength and $Fe_2O_3$ nanoparticle concentration (wt.%).